\documentclass[%
reprint,
amsmath,amssymb,
aps,
pra,
notitlepage
]{revtex4-2}% preamble:
\usepackage[bottom]{footmisc}
\usepackage[english]{babel}
\usepackage[utf8]{inputenc}
\usepackage{amsmath}
\usepackage{physics}
\usepackage{graphicx}
\graphicspath{{figures/}}
\usepackage[colorinlistoftodos]{todonotes}
\usepackage{lipsum}
\usepackage{multirow}
\usepackage{calc}
\usepackage[bottom]{footmisc}
\usepackage[]{hyperref}
\usepackage{booktabs}
\usepackage[normalem]{ulem}
\usepackage{float}
\usepackage{lineno}

\begin{document}

\title{Mid-Infrared Microscopy via Position Correlations of Undetected Photons}

\author{Inna Kviatkovsky$^{1,\star}$, Helen M. Chrzanowski$^{1}$, and Sven Ramelow$^{1,2}$}
\affiliation{$^{1}$Institut f\"ur Physik, Humboldt-Universit\"at zu Berlin, Berlin, Germany \\
$^{2}$IRIS Adlershof, Humboldt-Universit\"{a}t zu Berlin, Berlin, Germany \\
$^{\star}$Corresponding author: innakv@physik.hu-berlin.de}

\begin{abstract}
Quantum imaging with undetected photons (QIUP) has recently emerged as a new powerful imaging tool. Exploiting the spatial entanglement of photon pairs, it allows decoupling of the sensing and detection wavelengths, facilitating imaging in otherwise challenging spectral regions with mature silicon-based detection technology. All existing implementations of QIUP have so far utilised the momentum correlations within the biphoton state. Here, for the first time, we implement and examine theoretically and numerically the complementary scenario - utilising the tight position correlations formed within photon pair at birth. This image plane arrangement facilitates high resolution imaging with comparative experimental ease, and we experimentally show resolutions below 10$\mu$m at a sensing wavelength of 3.7 $\mu$m. Moreover, imaging a slice of mouse heart tissue at the mid-IR to reveal morphological features on the cellular level, we further demonstrate the viability of the technique for the life sciences. These results offer new perspectives on the capabilities of QIUP for label-free wide-field microscopy, enabling new real-world applications in biomedical as well as industrial imaging at inaccessible wavelengths.
\end{abstract}
\maketitle
The functionality to image samples in the mid-infrared (mid IR) offers a new perspective for problems of tremendous biological and industrial relevance. By exploiting the highly specific vibrational and rotational `fingerprints' of molecules as contrast mechanisms, one can obtain insights into the chemical and molecular structure inaccessible in traditional microscopy \cite{evans2008coherent,bhargava2012infrared,chrimes2013microfluidics}. The principle limitation, however, remains one of detection, with mid-IR imaging technology being prohibitively expensive while suffering from limited spectral response, spatial resolution or sensitivity. This absence of suitable detection options has lead to Raman-based imaging techniques \cite{cheng2015vibrational,muller2002imaging}, which require raster scanning, often making the technique too slow for many applications. Other approaches employ nonlinear wavelength conversion to the visible regime, where one can enjoy the comparable maturity of charge-coupled device (CCD) and complementary metal-oxide-semiconductor (CMOS) technology driven by the life sciences. Frequency up-conversion imaging shifts the detection frequency from the IR light to the desired visible while retaining the spatial and spectral information. This up-conversion imaging technique has been realised in the near- and mid-IR \cite{Dam2012,Demur2018,Junaid2019}.

An alternative approach and that which we consider here exploits quantum imaging with undetected photons (QIUP) \cite{lemos2014quantum}. Though its initial realisation was based on induced coherence without induced emission\cite{wang1991induced}, it can equivalently be described as an application of nonlinear interferometry \cite{chekhova2016nonlinear}. In QIUP, two nonlinear crystals are pumped sequentially and coherently with laser light, generating photon pairs through spontaneous parametric down conversation (SPDC). When the two processes are aligned such that any information distinguishing whether the biphoton was born on the first or second crystal is erased, the two processes interfere. The strength (visibility) and phase of this interference signature serve as the contrast mechanism for imaging - analogously so for similar applications in varied tasks including optical coherence tomography \cite{Valles:2018df,Paterova:2018kla,Vanselow:2020ia}, spectroscopy \cite{Kalashnikov:2016cl,Paterova:2018dz,Lindner20,Lindner21,Kutas21} and polarimetry\cite{Paterova19}. In contrast to ghost imaging schemes \cite{pittman1995optical,aspden2013epr,aspden2015photon}, in QIUP the detection of only one of the photons of the pair suffices to yield the information imprinted on the other. The strong spatial entanglement shared between the signal and idler pair, allows inference of the idler via the measurement of the signal \cite{HochrainerPNAS,Hochrainer:2017gh}, facilitating multi-mode (widefield) imaging. Crucially, this allows for the decoupling of the sensing and detection wavelengths when using a non-degenerate SPDC process\cite{lemos2014quantum}. An appealing way to use this advantage is to realise sensing at wavelengths for which advanced multi-pixel detection technologies are lacking. In this way, QIUP can shift the detection into the visible or near-IR wavelength ranges, where the far superior CMOS- and CCD-based sensor technologies operate.

This potential of QUIP for varied imaging applications, spanning wavelengths and applications \cite{Paterova:2020ei,Paterova:2021bi} motivates its relevance for the life sciences. Recently, QIUP was tailored to image at the microscopic length scale, so-called quantum microscopy with undetected photons (QMUP) \cite{kviatkovsky2020microscopy,paterova2020hyperspectral}. All prior realisations of QIUP and QMUP have exclusively exploited the momentum anti-correlations, imaging in the far-field of the crystal \cite{lemos2014quantum,HochrainerPNAS,Hochrainer:2017gh,Paterova:2020ei,kviatkovsky2020microscopy,paterova2020hyperspectral}. Here, we shift focus (both literately and figuratively) to instead examine imaging with the complementary position correlations that exist in the image plane \cite{viswanathan2021position}. We use these position correlations in the context of QIUP for the first time and demonstrate their advantage for microscopy in the mid-IR. In addition to demonstrating the viability of position correlations for imaging, we show that for microscopy, this approach dramatically simplifies the required optical overhead, obtaining sufficiently high resolutions with relative ease. 

\section*{Theory}

All existing QIUP and QMUP experimental implementations have utilised the momentum anti-correlations that arise due to transverse momentum conservation in the pair production process. However, the same spatial entanglement that gives rise to the aforementioned anti-correlations in momentum space, equally gives rise to strong correlations in the conjugate position space\cite{2020Fuenzalida}. Physically, these correlations stem from the tight position localisation created when the signal and idler pairs are born from the annihilation of a pump photon. Accordingly, the field of view (FoV) of the imaging system is then specified by the waist of the pump beam that illuminates the crystal, essentially defining an aperture within which the SPDC process can occur. For a Gaussian pump beam, the FoV accordingly is a Gaussian distribution with a full width at half maximum (FWHM) given by
\begin{equation}\label{FoV}
FOV_{IP}=\sqrt{2ln2}\frac{w_p}{M},    
\end{equation}
where $w_p$ is the pump waist at the crystal and $M$ is the optical magnification of the setup that scales the FoV after the crystal. 
The resolution, defined as the FWHM of the point-spread-function (PSF) of the system, is given by $res_{IP}=0.51\frac{\lambda_i}{NA_{lim}}$. The limiting numerical aperture $NA_{lim}$ is the minimum of the limit given by the optical components in the setup, or, more typically, given by the SPDC emission angle of the sensing (undetected) wavelength emitted from the crystal \cite{2020Fuenzalida}. For our system, the emission angle of the (undetected) idler light, scaled by subsequent magnification, is the dominant contribution. The resulting resolution in the image plane is
\begin{equation}\label{res}
res_{IP}=0.41\frac{\lambda_i}{\theta_{i}M}. 
\end{equation}
Here, $\theta_i$ is half the idler emission angle at FWHM, which relates to the crystal length through: $\theta_i=\lambda_i\sqrt{\frac{2.78}{\pi L}\frac{n_s n_i}{n_s \lambda_i+n_i\lambda_s}}$. One can equivalently and perhaps, more intuitively, consider the resolution to be a limitation arising from the thickness of the down-conversion crystal itself; the longer the crystal the more the ambiguity that arises regarding the birthplace of the signal and idler pair. The ratio of the FoV and resolution allows us to approximate the number of spatial modes per direction,
\begin{equation}\label{s.modes}
m_{IP}=\frac{\sqrt{2ln2}}{0.41}\frac{w_p\theta_i}{\lambda_i}\propto \frac{\omega_p}{\sqrt{L}}
\end{equation}
Optimisation of the imaging capabilities requires maximising the pump waist and minimising the crystal length, while also balancing the required illumination per mode. When inserting our experimentally determined values of $w_p=431 \pm 6 \mu m, \theta_i=0.0491 \pm 0.0003 \rm{rad}, \lambda_i=3.74 \pm 0.02 \mu m$  into Eqns.~\ref{FoV},\ref{res} and \ref{s.modes}, we obtain $FOV_{IP}=127 \pm 2 \mu m, res_{IP}=7.9 \pm 0.1 \mu m, m_{IP}=16.1 \pm 0.3$. By comparison, the corresponding far-field implementation \cite{kviatkovsky2020microscopy} theoretically predicts almost 50 \% more spatial modes per axis for the same pump light and crystal specifications. This is a consequence of some of the entanglement directly accessible in the far-field migrating into the imaginary part of the amplitude in the near-field, rendering it inaccessible with intensity measurements alone \cite{Reichert:2017dn,Just2013,Chan:2007ds}. Despite this reduction in available spatial modes, here we show that image plane imaging offers an advantage: imaging at microscopic length scales with large reduction in optical complexity. Requiring less magnification, fewer optical elements and consequently shorter interferometric arms, the demonstrated resolution in the image plane is thrice superior to that of achieved in the far field \cite{kviatkovsky2020microscopy}. 

\section*{Experiment}
\begin{figure}[h!]
\centering\includegraphics[width=\columnwidth]{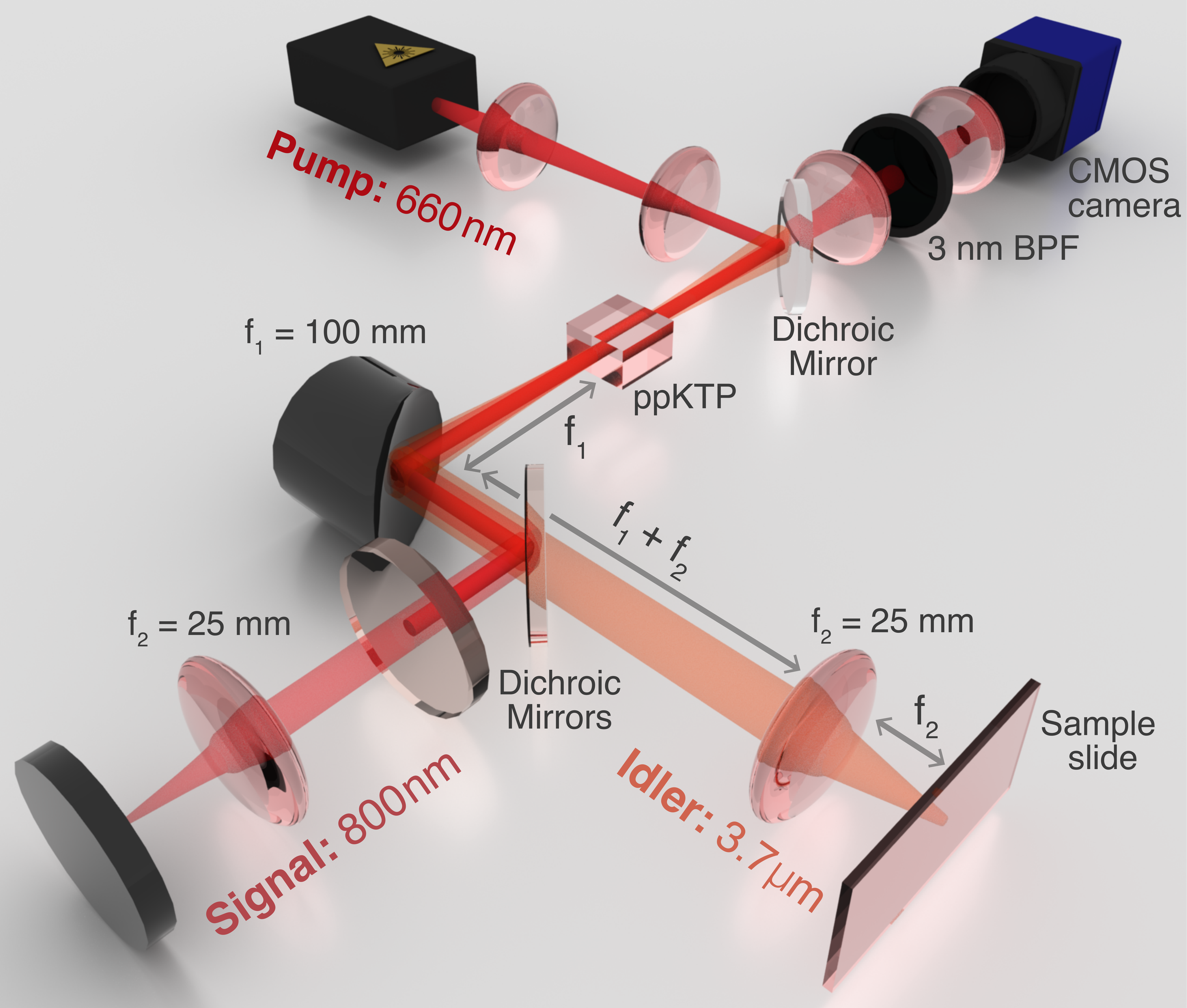}
\caption{Experimental Setup: A continuous wave pump light at 660 nm stimulates a highly non-degenerate, collinear SPDC process in a Michelson-type interferometer. The signal (800 nm) and idler (3.7 $\mu$m) fields generated on the first pass are split via a dichroic mirror, allowing the idler to probe the sample, before being recombined and travelling collinear with the coherent pump field back into the crystal. The pump is independently reflected back via a separate cold mirror in the signal arm. The signal field emerging after the second pass of the crystal is spectrally filtered via a band-pass filter (BPF) and then imaged onto a CMOS camera, revealing the spatial information obtained by the idler when probing the sample.}
\label{setup}
\end{figure}

The experimental setup is detailed in Fig.~\ref{setup} and exploits a Michelson-type configuration to realise the nonlinear interferometer. A 660 nm CW pump laser illuminates the ppKTP crystal, with a 431 \textmu m waist maximally covering the crystal aperture. The ppKTP crystal is quasi-phase matched for a collinear type-0 process and specifically engineered\cite{vanselow2019ultra} to produce simultaneously highly non-degenerate and broadband photon pairs. The broadband SPDC emission is due to the group velocity matching of the signal and idler, with spectral widths of 780-830 nm and 3.4-4.3 $\mu$m respectively at room temperature \cite{vanselow2019ultra}. After the crystal, an off-axis parabolic mirror (OPM) is placed at its focal distance for achromatic collimation (focusing) of the emerging (returning) signal and idler. Using a dichroic mirror (DM), the idler is then split from the signal and pump, with the pump subsequently back-reflected using a cold mirror to preserve the desired imaging condition. The idler and signal fields are then each focused to align the image plane of the crystal with their respective end mirrors. A sample for imaging is placed on the low-e slide that serves as an end mirror for the idler arm. The reflected idler and signal fields then back-propagate and are focused into the crystal, interfering with signal and idler fields generated upon the second pass of the pump through the crystal. The idler and pump light emerging from the second pass of the crystal are then discarded and the signal light is imaged onto a CMOS camera. Prior to detection, the signal field is further filtered using a band-pass filter (3.5 nm FWHM) and a telescopic arrangement is used to position the CMOS camera in the image plane of the crystal. 

It is crucial to carefully align the setup to ensure optimal spatial overlap between the biphoton fields generated in the two passes through the crystal. This ensures indistinguishably that will be manifested in the visibility of the interferometric image captured. This requirement is achieved by simultaneously matching the interferometric arms within the coherence length of the detected signal light and carefully tuning the optical components to fulfil the imaging conditions. Any deviation from those requirements will result in departure from optimal visibility and potentially, resolution, of the imaging system.

\section*{Results}

\begin{figure}[h!]
\centering\includegraphics[width=\columnwidth]{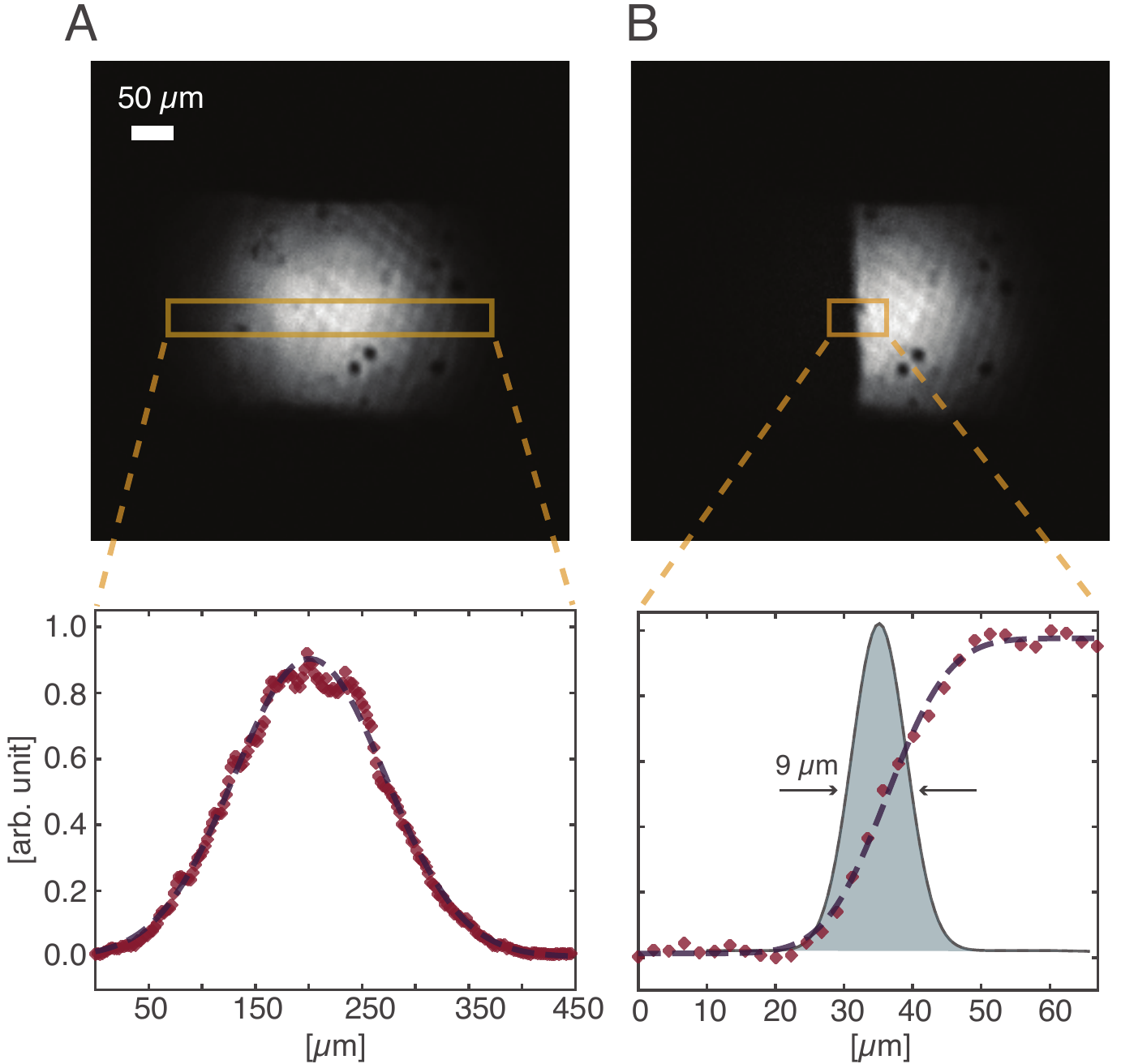}
\caption{Characterization of the imaging arrangements. (A) FoV. The measured data (pink points) was fitted with a Gaussian (purple line) yielding a FoV of $161\mu m$ (FWHM). (B) Edge response in the top row, the measured data (pink points) was fitted with an error function (purple line), differentiation of the error function gives a Gaussian shaped (marked in blue) point spread function (PSF). The resolution, determined by the FWHM of the PSF is $9\mu m$. }
\label{characterisation}
\end{figure}

Fig.~\ref{characterisation} presents the characterisation of our imaging system at a magnification of $M=4$. The resolution, obtained via an edge knife response, is $9\pm 1\mu m$ (FWHM) and the obtained FoV is $161 \pm1\mu m$  (FWHM). These values compare favourably to the theoretical values for the resolution and FoV of $7.8 \mu m$ and $127 \mu m $ respectively, albeit at a lower than anticipated magnification, as the corresponding number of spatial modes -- $18\pm 2$ per axis -- is in agreement with the theoretical value. 

The experimental results are summarised in Table 1. We attribute the systematic deviation of the larger obtained FoV and resolution to a reduced magnification realised in practice for the 4f system, when compared to the anticipated magnification of exactly 4.  As predicted by the theory, QMUP via position correlations allows one to access high resolutions with a simplified optical system, notably when compared to realisations in the Fourier plane \cite{kviatkovsky2020microscopy}. This is an inherent characteristic of Fourier plane imaging due to the divergence of the biphoton field at the crystal exit, resulting in a large illumination spot (FoV) in the far field.

Due to the interferometric nature of the imaging technique, a distortion of the biphoton wavefront can result in ambiguities; for a single image it can be ambiguous whether a dark region references absorption, or rather, destructive interference. This is particularly relevant for biological or industrial applications, where complex morphologies underpin the imaging motivation. By scanning axially within the coherence length, one can obtain a pixel-wise visibility of the sample, allowing pixel-wise reconstruction of both the transmissivity and the phase. This has a secondary advantage of increasing the `effective' FoV of the image; in contrast to the intensity distribution of the illumination spot itself, the visibility distribution is considerably flatter. The effective number of spatial modes in the visibility images is here approximately double, resulting in roughly 900 spatial modes for the 2D wide field imaging arrangement.

\begin{figure}[h!]
\centering\includegraphics[width=\columnwidth]{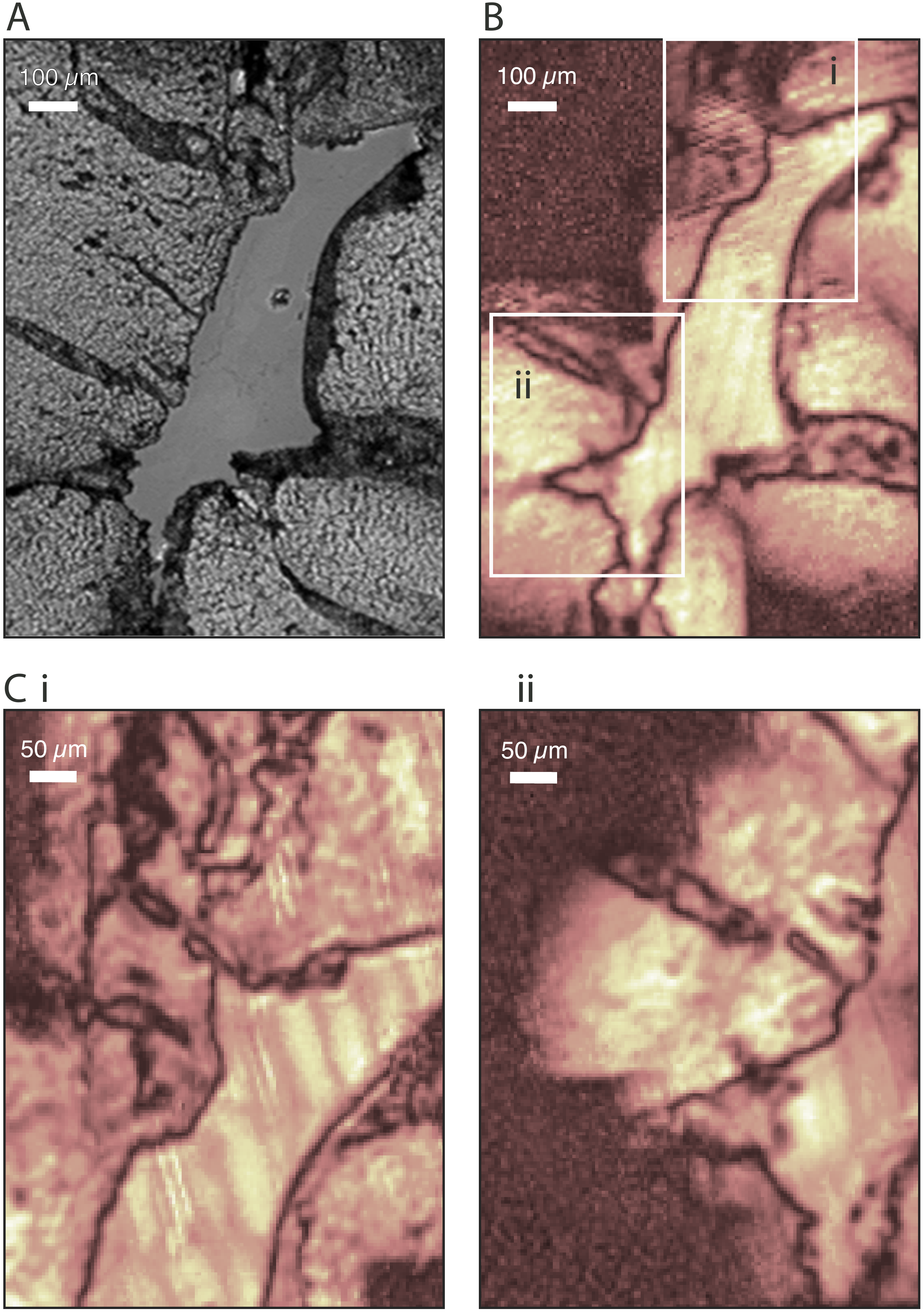}
\caption{Histology sample of a mouse heart imaged with (A) bright field microscopy with visible light for illustration which part of the sample we investigated with our method. (B) Mid-IR microscopy of the same sample with undetected photons for absorption imaging with a 2-fold magnification. (C) Higher resolution absorption images, taken with the a 4-fold magnification arrangement. Images are formed by stitching roughly 7 wide-field absorption images (translating the sample transversely), each absorption image  reconstructed by averaging 6 images at 1 s integration time for 6 axial positions within the coherence length of the biphoton (longitudinal scan).}
\label{biosample}
\end{figure}

Using this scanning technique, a thin unstained slice of a mouse heart tissue was mounted on a low-e slide and imaged. Complementary to the above characterised imaging system (M=4), a lower magnification (M=2) configuration was also used to acquire a larger scale absorption image (Fig.~\ref{biosample}(B)). To enable comparison, a bright field image acquired with a standard visible microscope is presented in Fig.~\ref{biosample}(A). The left ventricle of the mouse heart and surrounding structure is visible in the images, revealing various morphological features. In Fig.~\ref{biosample}(c) two smaller regions in the sample were characterised using the larger magnification arrangement (M=4), with this increased resolution revealing additional features that are not visible in the lower magnification images. The additional morphological information revealed in the mid-IR image is inaccessible in the bright field image, which can be attributed to the considerably reduced scattering at mid-IR wavelengths.

\section*{Discussion}

\begin{table}[h!]
\centering
\begin{tabular}{||c c c||} 
 \hline
 & Experiment & Theory \\ [0.5ex] 
 \hline\hline
 FoV ($\mu$ m) & $161 \pm 1$ & $127 \pm 2$ \\ 
 Resolution ($\mu$ m) & $9\pm 1$ & $7.9\pm 0.1$ \\
 Spatial modes & $18\pm2$ & $16.1 \pm 0.3$ \\
  [1ex] 
 \hline
\end{tabular}
\caption{QMUP in the image plane - experiment vs. theory}
\label{table:1}
\end{table}

One technical disadvantage of QMUP in the image plane is the reduced homogeneity of the illumination distribution. In the image plane, we illuminate with the photon birth zone itself, revealing inhomogeneities arising from small defects in the crystal, dust or imperfect pump modes. By contrast, the Fourier plane provides a very homogeneous illumination distribution, its uniformity being a consequence of the smoothness of the phase-matching curve itself. This disadvantage is reminiscent of classical microscopy, where schemes usually avoid illumination in the image plane of the light source to circumvent the analogous noise contributions \cite{mertz2019introduction}. Here, where the correlations between the planes are indispensable, it cannot be easily avoided, but can be eliminated with care towards the crystal and pump. The higher noise in the image plane is one potential contributor to the reduction of visibility, as the background noise in the image plane is inverted on the second pass (and thus does not cancel out). A second source of potential visibility reduction stems from the high magnification, which results in a lower depth of focus and thus a higher sensitivity to alignment errors. Both of these sources of reduced visibility are technical in nature and not fundamental limitations.

In contrast to our prior work in the Fourier plane microscopy, we suffer no degradation from the theoretical performance of our system despite imaging at considerably smaller length scales. While imperfect matching of the image plane at the sample and camera will likely degrade the resolution of the imaging system, the short depth of focus aids in matching these conditions very precisely. The 9 $\mu$m resolution presented here was realised in a simple 4f setup with a standard aspheric lens (f = +25 mm, Thorlabs) serving as the magnifying lens. This basic 4f arrangement affords superior compactness and stability, and with targeted optical engineering, should attain diffraction-limited resolutions at the state-of-the-art of mid-IR imaging.

\begin{figure}[h!]
\centering\includegraphics[width=\columnwidth]{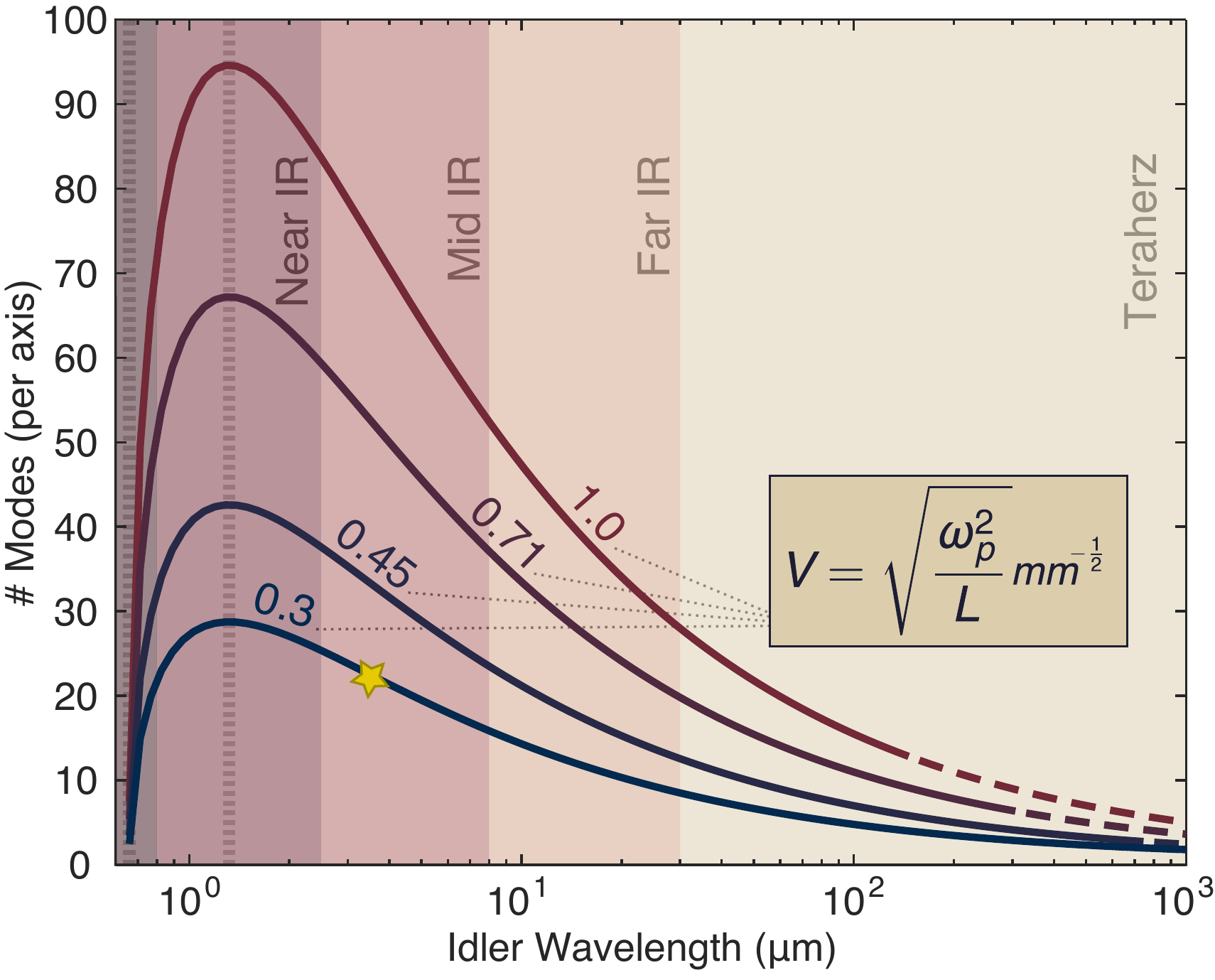}
\caption{Numerical simulation of the number of spatial modes available for wide field imaging (as characterised by the Schmidt number) as a function of increasing idler wavelength for four different `effective' crystal apertures, V. For instance, the four decreasing values of V (= 1.0, 0.71, 0.45 and 0.3 mm$^{-1/2})$ correspond to increasing values of crystal length, L (= 1, 2, 4 and 11 mm) for a pump waist of $\omega_p$ = 1 mm. The pump wavelength of the pump light is fixed at 660 nm and for simplicity the refractive indices are assumed to be 1.5 across the examined range. The pump wavelength and degeneracy point (1.32 $\mu$m) are highlighted by vertical dashed lines. The dashing of curves lines indicates a parameter regime where these results, derived under the paraxial approximation, may no longer faithfully describe the system. The imaging capacity of the system presented here ($\omega_p$ = 430 $\mu$m and L = 2 mm) is indicated by a star.}
\label{farIR}
\end{figure}

The clear success of this approach for imaging and microscopy in the mid-IR motivates the question of its limitations - can we envisage this approach to wide field imaging stretching beyond its near and mid-IR, towards the far IR and even terahertz wavelengths\cite{Kutas21}, where detection technologies are even more limited? Fig. \ref{farIR} presents a theoretical analysis of this question in terms of available spatial modes. It shows a decrease in wide field imaging capacity QIUP with increasing (undetected) wavelength. The increasing non-degeneracy between the signal and idler energy that enables `silicon imaging' at silicon incompatible wavelengths, must be traded off against a decrease in the available spatial entanglement and thus the wide field imaging capacity. Wide field QIUP applications remain promising into the far IR, but become increasingly unfavourable as we approach the terahertz regime. Furthermore, this analysis does not consider factors including material absorption, parasitic seeding and the necessity for long crystals that complicate the imaging at very long wavelengths. This analysis, however, does not preclude single-pixel based scanning approaches to imaging tasks at these wavelengths.

The approach presented here also opens up the possibility of an imaging regime previously inaccessible at far wavelengths: facilitating shot-noise limited imaging with high quantum efficiencies at exceptionally low illumination powers. This is the consequence of the intrinsic (theoretical unity) efficiency of the nonlinear interferometer in the low-gain regime, where any (mid-IR) idler photon has its (silicon-compatible) partner. Accordingly, any imaging information carried by that idler photon can be transferred perfectly to the signal photon. Therefore, in the absence of additional loss and mode mismatch, the noise performance of the mid-IR imaging is determined by the properties of the silicon camera, where shot noise–limited images are accessible with only a few 1000 of photons per pixel per second or less. Here, the images were obtained at mid-IR illumination levels of only a 15 pW -- with the light within our current detection bandwidth amounting to only 2-3 pW. Such low illuminations are significant for the understanding of photoreceptive samples, where the sensing illumination itself can invariably interfere with the cellular and molecular mechanisms one seeks to understand. 

\section*{Conclusion}

In conclusion, we have presented the first experimental realisation of QIUP via position correlations. We have shown that imaging with position correlations presents advantages over its predecessors for the task of microscopy, allowing access to resolutions below 10 $\mu m$ at a sensing wavelength of 3.7 $\mu m$ with no observable deviation from theoretical predictions. This improved resolution permitted mid-IR imaging of an unstained tissue from a mouse heart with several orders of magnitude less light than any comparative methods. The presented results further extend the growing toolbox of QMUP and take us another step closes toward real-world applications.

\subsection*{Method}

Slide Preparation: 9-12 weeks old C57BL/6J mice were sacrificed by cervical dislocation. Hearts were removed, rinsed in ice-cold saline and placed in 4 \% formalin. After 48
to 72 hrs fixation the tissue was rinsed with (Phosphate-buffered saline) PBS and then embedded
in paraffin. The paraffin-embedded hearts were cut in transverse sections to a thickness of 2 to 3 µm
and transferred to a low-e slide.

\subsection*{acknowledgements}
The authors want to acknowledge Ellen G. Avery and Hendrik Bartolomaeus for providing the bio-samples and helping with the interpretation of the results. The authors also thank Sergey Berezinski for assistance with preparation of the figures. This work was funded by Deutsche Forschungsgemeinschaft (RA 2842/1-1).

%\bibliography{references}


\end{document}